# Intrusion Detection in Mobile Ad Hoc Networks Using Classification Algorithms

Aikaterini Mitrokotsa, Manolis Tsagkaris and Christos Douligeris

**Abstract** In this paper we present the design and evaluation of intrusion detection models for MANETs using supervised classification algorithms. Specifically, we evaluate the performance of the MultiLayer Perceptron (MLP), the Linear classifier, the Gaussian Mixture Model (GMM), the Naïve Bayes classifier and the Support Vector Machine (SVM). The performance of the classification algorithms is evaluated under different traffic conditions and mobility patterns for the Black Hole, Forging, Packet Dropping, and Flooding attacks. The results indicate that Support Vector Machines exhibit high accuracy for almost all simulated attacks and that Packet Dropping is the hardest attack to detect.

## 1 Introduction

The adoption of Mobile Ad hoc networks (MANETs) has increased in recent years mainly due to their important advantages and their broad applicability. MANETs can be defined as dynamic peer-to-peer networks that consist of a collection of mobile nodes. The nodes employ multi-hop information transfer without requiring an existing infrastructure. Although MANETs are characterized by great flexibility and are employed in a broad range of applications, they also present many inherent vulnerabilities that increase their security risks. Due to their dynamic and cooperative nature, MANETs demand efficient and effective security mechanisms in order to be safeguarded. Intrusion prevention can be used as a first line of defense in order to reduce possible intrusions but undoubtedly, it cannot eliminate them. Intrusion de-

Aikaterini Mitrokotsa
Department of Computer Science, Vrije Universiteit, De Boelelaan 1081a, 1081HV Amsterdam Netherlands, e-mail: katerina@few.vu.nl

Manolis Tsagkaris, Christos Douligeris
Department of Informatics, University of Piraeus, Karaoli and Dimitriou 80, 18534 Piraeus, Greece, e-mail: m_tsag@yahoo.com, cdoulig@unipi.gr

tection using classification algorithms can help us to effectively discriminate "normal" from "abnormal" behaviour and thus, detect possible intrusions. Therefore, intrusion detection, serving as a second line of defense, is an indispensable part of reliable communication in MANETs.

Intrusion Detection has a long history of research in wired network defense but it is still in its infancy in the area of wireless ad hoc networks. There is though, a small number of proposed Intrusion Detection Systems (IDS) for wireless ad hoc networks. Zhang and Lee [14] proposed the first (high-level) IDS approach specific for ad hoc networks. They proposed a distributed and cooperative anomaly-based IDS, which provides an efficient guide for the design of IDS in wireless ad hoc networks. They focused on an anomaly detection approach based on routing updates on the Media Access Control (MAC) layer and on the mobile application layer.

Huang and Lee [6] extended their previous work by proposing a cluster-based IDS, in order to combat the resource constraints that MANETs face. They use a set of statistical features that can be derived from routing tables and they apply the classification decision tree induction algorithm C 4.5 in order to detect "normal" vs. "abnormal" behavior. The proposed system is able to identify the source of the attack, if the identified attack occurs within one-hop.

Deng et al. [2] proposed a hierarchically distributed and a completely distributed intrusion detection approach. The intrusion detection approach used in both of these architectures focuses on the network layer and it is based on a Support Vector Machines (SVM) classification algorithm. They use a set of parameters derived from the network layer and suggest that a hierarchically distributed approach may be a more promising solution versus a completely distributed intrusion detection approach. Liu et al. [8] proposed a completely distributed anomaly detection approach. They investigated the use of the MAC layer in order to profile normal behavior of mobile nodes and then apply cross-feature analysis [5] on feature vectors constructed from the training data.

Although some use of classification algorithms was present in all of the aforementioned papers, almost none contained comparisons between methods, apart from [14]. Thus, there is a lack of evidence to support the use of one algorithm compared to others, when it comes to intrusion detection in MANETs. Furthermore, there is virtually no data on the performance of such algorithm under different traffic conditions (i.e. mobility, number of malicious nodes), and how such meta-algorithmic parameters such as the sampling interval should be selected. The selection of the sampling interval is particularly important, as there could be a trade-off between good classification performance and quick response.

The main novelty of this paper relative to the aforementioned work, lies in the following. Firstly, we are performing a comparison between multiple, well-known, classification models, using labeled training data. Secondly, this comparison is done in a principled experiment, where the hyperparameters that must be tuned for each classification model are selected using a specific procedure, which is the same for all models. In this way we are sure of a fair comparison, as when employed in the field, algorithms would have to be tuned before seeing any actual test data. Finally, we examine how the performance of each algorithm changes for various values of network mobility, sampling interval and the number of malicious nodes. In our previous work [9], we have used a neural network in order to classify normal traffic and selective packet dropping attack. Our goal is to explore whether there exists a classification algorithm that demonstrates superior performance in detecting a given attack category for all, or most traffic conditions. Furthermore, we examine the importance of the sampling interval time of the statistical features and consequently how quickly the intrusion detection can be performed.

More specifically, we present a performance comparison of five efficient and commonly used classification algorithms (MultiLayer Perceptron (MLP), Linear classifier, Gaussian Mixture Model (GMM), Naïve Bayes classifier and Support Vector Machines (SVM)) applied to intrusion detection in MANETs. We use features from the network layer and evaluate the performance of the classification algorithms for the detection of the *Black hole*, *Forging*, *Packet Dropping* and *Flooding* attacks.

Following this introduction, the paper is organized as follows. Section 2 describes the quality metrics used for the classification comparison as well as the classification models employed. Section 3 describes the experiments that have been performed, the experimental setup and the results. Section 4 concludes the paper and discusses some future work.

## 2 Intrusion Detection Using Classification

We employ classification algorithms in order to perform intrusion detection in MANETs. Compared to other methods, classification algorithms have the advantage that they are largely automated and that they can be quite accurate. They have extended applications including intrusion detection in wired networks [7], great literature coverage and extended experimental use that denote their efficiency.

### 2.1 Intrusion Detection Model

The IDS architecture we adopt is composed of multiple local IDS agents, that are responsible for detecting possible intrusions locally. The collection of all the independent IDS agents forms the IDS system for the MANET. Each local IDS agent is composed of the following components:

*Data Collector*: is responsible for selecting local audit data and activity logs.

*Intrusion Detection Engine*: is responsible for detecting local intrusions using local audit data. The local intrusion detection is performed using a classification algorithm. Firstly, it performs the appropriate transformations on the selected labeled audit data. Then, it computes the classifier using training data and finally applies the classifier to test local audit data in order to classify it as "normal" or "abnormal".

*Response Engine*: If an intrusion is detected by the Detection Engine then the Response Engine is activated. The Response Engine is responsible for sending a local and a global alarm in order to notify the nodes of the mobile ad hoc network about the incident of intrusion.

## 2.2 Algorithmic Comparisons and Quality Metrics

When comparisons are made between algorithms, it is important to use the same measure of quality. For a given classification algorithm $f : \mathscr{X} \to \mathscr{Y}$, where $\mathscr{X}$ is the observation space and $\mathscr{Y}$ is the set of classes, a common measure of quality is the classification error C measured over an independent test set D,

$$\hat{E}(C|D) = \frac{1}{|D|} \sum_{d \in D} C(f(x_d), y_d), \qquad (1)$$

where $x_d$ is the observation of example $d$ and $y_d$ is its class and $C(y', y) = 0$ when $y = y'$ and 1 otherwise. However, it is important to note that in most of the literature, the *Detection Rate* (*DR*) and the *False Alarm* (*FA*) rate are used instead:

$$DR = \frac{TP}{TP + FN}, \qquad FA = \frac{FP}{TN + FP} \qquad (2)$$

where $TP$, $TN$, $FP$, $FN$, denote the number of true ($TP$ & $TN$) and false ($FP$ & $FN$) positives and negatives respectively. The goal of an effective intrusion detection approach is to reduce to the largest extent possible the *False Alarm* rate (*FA*) and at the same time to increase the *Detection Rate* (*DR*).

## 2.3 Classification Models

In this section we describe the classification models we have used in order to perform intrusion detection i.e., the MultiLayer Perceptron (MLP), the Linear model, the Gaussian Mixture model (GMM), the Naïve Bayes model and the SVM model. All these models require labelled training data for their creation.

A specific instance of an MLP can be viewed simply as a function $g : \mathscr{X} \to \mathscr{Y}$, where $g$ can be further defined as a composition of other functions $z_i : \mathscr{X} \to \mathscr{Z}$. In most cases of interest, this decomposition can be written as $g(x) = Kw'z(x)$ with $x \in X$, $w$ being a parameter vector, while $K$ is a particular kernel and the function $z(x) = [z_1(x), z_2(x), ...]$ is referred to as the *hidden layer*. For each of those, we have $z_i(x) = K_i(v_i'x)$ where each $v_i$ is a parameter vector, $V = [v_1, v_2, ...]$ is the parameter matrix of the hidden layer and finally $K_i$ is an arbitrary kernel. For this particular application we wish to use an MLP $m$, as a model for the conditional class probability given the observations, i.e.

$$P(Y = y | X = x, M = m), \qquad y = g(x). \tag{3}$$

The case where there is no hidden layer is equivalent to $z_i = x_i$, which corresponds to the *Linear model*, the second model into consideration.

The GMM, the third model under consideration, will be used to model the conditional observation density for each class, i.e. $P(X = x | Y = y, M = m)$.

This can be achieved simply by using a separate set of mixtures $U_y$ for modeling the observation density of each class $y$. Then, for a given class $y$ the density at each point $x$ is calculated by marginalizing over the mixture components $u \in U_y$, for the class, dropping the dependency on $m$ for simplicity:

$$P(X = x | Y = y) = \sum_u P(X = x | U = u) P(U = u | Y = y). \tag{4}$$

Note that the likelihood function $P(X = x | U = u)$ will have a Gaussian form, with parameters the covariance matrix $\Sigma_u$ and the mean vector $\mu_u$. The term $P(U = u | Y = y)$ will be represented by another parameter, the component weight. Finally, we must separately estimate P(Y=y) from the data, thus obtaining the conditional probability given the observations:

$$P(Y = y | X = x) = \frac{1}{Z} P(X = x | Y = y) P(Y = y), \tag{5}$$

where $Z = \sum_{y \in Y} P(X = x | Y = y) P(Y = y)$ does not depend on y and where we have again dropped the dependency on m.

The fourth model under consideration is the Naïve Bayes model which can be derived from the Gaussian Mixture Model (GMM) when there is only one Gaussian mixture.

The last model we evaluated in order to perform intrusion detection in MANETs is the Support Vector Machine (SVM) [1] model, which uses Lagrangian methods to minimise a regularized function of the empirical classification error. The SVM algorithm finds a linear hyperplane separation with a maximal margin in this hyperspace. The points that are lying on the margin are called support vectors. The main parameter of the algorithm is $c$, which represents the trade-off between the size of the margin and the number of violated constraints, and the kernel $K(x_i, x_j)$. In this work we will utilize SVMs with a gaussian kernel of the form $K(x_i, x_j) = \frac{1}{\sqrt{2\pi}\sigma} \exp(-\|x_i - x_j\|^2 / \sigma^2)$.

## 3 Experiments

In order to examine the performance of the classification algorithms, we conducted a series of experiments under varying conditions. In our experiments we performed comparisons in terms of the classification cost defined in equation (1) using ten different models: MLP, Linear, GMM with diagonal covariance matrices, Naïve Bayes (GMM with a single Gaussian) and SVM models for binary and multiclass classification. In all cases we use the same set of features, as explained in the following section. In binary classification all attacks are lumped together so the task is just to identify the presence of an attack. The multiclass task requires the correct identification of each attack type. We investigated which sampling interval for statistical features is more appropriate, which algorithm presents the best performance in terms of *Detection Rate* (*DR*) and *False Alarm* (*FA*) rate and which algorithm is better for the detection of specific attacks. Furthermore, we investigated how the performance of the classification algorithms change when we vary the number of malicious nodes in the network and when we vary the mobility of the network.

## *3.1 Simulation Environment*

In order to evaluate our approach we simulated a mobile ad hoc network (MANET) and we conducted a series of experiments. For our experiments we have made the assumption that the network has no preexisting infrastructure and that the employed ad hoc routing protocol is the Ad hoc On Demand Distance Vector (AODV). We implemented the simulator within the GloMoSim [4] library. Our simulation models a network of 50 hosts placed randomly within an 850 x 850 $m^2$ area. Each node has a radio propagation range of 250 meters and the channel capacity was 2 Mbps. The nodes in the simulation move according to the 'random way point' model. At the start of the simulation, each node waits for a pause time, then randomly selects and moves towards a destination with a speed uniformly lying between zero and the maximum speed. On reaching this destination it pauses again and repeats the above procedure till the end of the simulation. The minimum and maximum speed is set to 0 and 20 m/s, respectively, and pause times at 0, 200, 400, 700 sec. The simulation time of the experiments was 700 sec, thus a pause time of 0 sec corresponds to the continuous motion of the node and a pause time of 700 sec corresponds to the time that the node is stationary.

Each node is simulated to generate Constant Bit Rate (CBR) network traffic. The size of the packets sent by each node varies from 128 to 1024 bytes. We have studied the performance of the classification algorithms for various sampling intervals (5, 10, 15, 30 sec) in order to study how quickly these algorithms can perform intrusion detection. The sampling interval dictates both the interval for which the statistical features are calculated, and the period between each classification decision. We expect that longer intervals may provide more information, but with the cost of slower detection. We have also evaluated the performance of the classification algorithms for 5, 15 and 25 malicious nodes. In each case the number of all nodes in the network is set to 50.

In our experiments, we have simulated four different types of attacks:

- **Flooding attack:** We have simulated a flooding attack [13] for multiple paths in the network layer, where each malicious node sends forged RREQ (Route REQuest) packets randomly to all nodes of the network every 100 msec.

- **Forging attack:** We have simulated a forging attack [12] for RERR (RouteERRor) packets, where each malicious node modifies and broadcasts (to a selected victim) a RERR packet every 100 msec leading to repeated link failures.
- **Packet Dropping attack:** We have simulated a selective packet dropping [3] attack, where each malicious node drops all RERR packets leading legitimate nodes to forward packets in broken links.
- **Black Hole attack:** In a black hole attack [10], a malicious node advertises spurious routing information, thus receiving packets without forwarding them but dropping them. In the black hole attack we have simulated the scenario where each time a malicious-black hole node receives a RREQ packet it sends a RREP (RouteREPly) packet to the destination without checking if the node has a path towards the selected destination. Thus, the black hole node is always the first node that responds to a RREQ packet and it drops the received RREQ packets. Furthermore, the malicious-black hole node drops all RREP and data packets it receives if the packets are destined for other nodes.

A very important decision to be made is the selection of feature vectors that will be used in the classification. The selected features should be able to represent the network activity and increase the contrast between "normal" and "abnormal" network activity. We have selected the following features from the network layer:

- *RREQ Sent:* indicates the number of RREQ packets that each node sends.
- *RREQ Received:* indicates the number of RREQ packets that each node receives.
- *RREP Sent:* indicates the number of RREP packets that each node receives.
- *RError Sent:* indicates the number of RError (Route Error) packets that each node receives.
- *RError Received:* indicates the number of RError packets that each node sends.
- *Number of Neighbors:* indicates the number of one-hop neighbors that each node has.
- *PCR (Percentage of the Change in Route entries):* indicates the percentage of the changed routed entries in the routing table of each node. PCR is given by $(|S_2 - S_1| + |S_1 - S_2|)/|S_1|)$, where $(S_2 - S_1)$ indicates the newly increased routing entries and $(S_1 - S_2)$ indicates the deleted routing entries during the time interval $(t_2 - t_1)$.
- *PCH (Percentage of the Change in number of Hop):* indicates the percentage of the changes of the sum of hops of all routing entries for each node. PCH [11] is given by $(H_2 - H_1)/H_1$, where $(H_2 - H_1)$ indicates the changes of the sum of hops of all routing entries during the time interval $(t_2 - t_1)$.

For each sampling interval time (5, 10, 15, 30 sec) we have created one training dataset, where each training instance contains summary statistics of network activity for the specified interval using all the above features and in addition, the type of attack performed during this interval. This enables us to use supervised learning techniques for clasification. Each training dataset was created by running different simulations with duration 700 sec for different network mobility (pause time equal to 0, 200, 400, 700 sec) and varying numbers of malicious nodes. The derived datasets from each of these simulations were merged and one training dataset was produced for each sampling interval. A similar procedure was followed in order to produce the testing datasets.

## *3.2 Algorithmic Technical Details*

In order to select the best parameters for each algorithm we performed a 10-fold cross validation [15] on the training dataset, which were created with random sampling. For each of the 10 folds, we selected 1/10th of the dataset for evaluation and the remaining for training. We then used the parameters selected to train a new model using all of the training set for each algorithm, which was the model that was evaluated.

For the MLP we tuned three parameters, i.e. the *learning rate ($\eta$)* and the *number of iterations (T)* used in the stochastic gradient descent optimization as well as the *number of hidden units (nh)*. Keeping $nh$ equal to 0 we selected the appropriate $\eta$ among values that range between 0.0001 and 0.1 with step 0.1 and the appropriate $T$ selecting among 10, 100, 500 and 1000. Having selected the appropriate $\eta$ and the appropriate $T$, we examined various values in order to select the appropriate $nh$. We selected the best among 10, 20, 40, 60, 80, 100, 120, 140, 160, 320. Additionally, we used the MLP model with no hidden units as a Linear model.

For the GMM we also tuned three parameters, i.e. the *threshold ($\theta$)*, the *number of iterations (T)* and the *number of Gaussian Mixtures (ng)*. Keeping stable $ng$ (equal to 20) we selected the appropriate $\theta$ among values that range between 0.0001 and 0.1 with step 0.1 and the appropriate $T$ among 25, 100, 500 and 1000. For the selection of the appropriate $ng$, after selecting the appropriate $\theta$ and the appropriate $T$, we examined various values for the $ng$ and the selected best among 10, 20, 40, 60, 80, 100, 120, 140, 160, 320. Additionally, we used the GMM model with one Mixture component as a Naïve Bayes model.

For the SVM we tuned two parameters, i.e. the standard deviation ($\sigma$) for the gaussian kernel and the regularisation parameter $c$ which represents the trade-off between the size of the margin and the number of misclassified examples. For the selection of the appropriate combination of $\sigma$ and $c$, we examined various values for the $\sigma$ (1, 10, 100, 1000) and the $c$ (1, 10, 100, 1000) and selected the best.

## *3.3 Experimental Results*

First, we examined which is the most appropriate sampling interval time of the used statistical features. We used the models produced by the training datasets with the appropriate parameters (Sect. 4.2). Figure 1 depicts the average as well as the minimum and the maximum classification error of the testing datasets. While 15 sec appears to be the best sampling interval in the test dataset on average, this does not hold for all cases. It will probably be best to use different intervals for different attacks and classification algorithms, but this is beyond the scope of the current work.

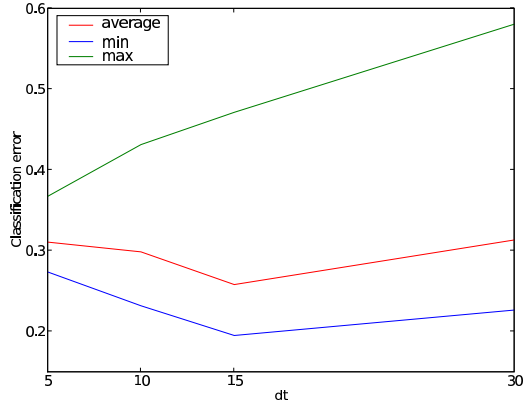

**Fig. 1** Classification error versus sampling interval (dt)

Figure 2a depicts the average *Detecton Rate* (DR) and the *False Alarm* (FA) rate for all classification algorithms both for binary and multiclass classification. The best *Detection Rate* (DR) is achieved for the MLP classifier for multiclass classification and is equal to 78.95%, while the corresponding *False Alarm* (FA) rate is equal to 12.92%. The second best classifier with a high *Detection Rate* (DR) equal to 77% is achieved with the SVM classifier for multiclass classification. The corresponding *False Alarm* (FA) rate is quite lower compared to the one achieved with the MLP classifier and is equal to 0.97%. The classifier that presents the poorest performance is the Naïve Bayes classifier with *Detection Rate* (DR) equal to 41.88% and *False Alarm* (FA) rate equal to 0.47%.

Figure 2b depicts the *Detection Rate* (DR) for each type of attacks (Black hole, Forging, Packet Dropping, Flooding), for all classification models. It is obvious that for all classifiers the best *Detection Rate* (DR) is achieved for the *Flooding* attack, while the most difficult attack to detect is the *Packet Dropping* attack. The best *Detection Rate* (DR) for the *Black hole* and the *Flooding* attack is achieved with the *Linear* classifier for multiclass classification and is equal to 87.75% and 96.06% correspondingly. The best *Detection Rate* (DR) equal to 76.86% and 73.88% correspondingly for the *Packet Dropping* and *Forging* attack is achieved again with the *Linear* classifier but for binary classification.

Figure 3 depicts the average, minimum and maximum *classification error* for all classification algorithms for binary and multiclass classification versus the number of malicious nodes (Fig. 4a and Fig. 5a) that exist in the network and the mobility (pause time) (Fig.4b, Fig. 5b) of the network. In order to investigate the performance of the classification algorithms versus the number of malicious nodes the testing datasets were generated by keeping the mobility of the network stable (pause time equal to 200 sec) and by changing the number of malicious nodes that exist in the network (5, 15, 25). Similarly, in order to investigate the performance of the classification algorithms versus the mobility of the network, we have kept stable the number of malicious nodes in the network (equal to 15) and changed the mobility of the network (pause time equal to 0, 200, 400, 700 sec). It is clear that in both (Fig. 3a and Fig. 3c) the binary and the multiclass classification, the higher the number of malicious nodes in the network, the easier it is to detect possible intrusions. Furthermore, it is clear that in both cases (Fig. 3b, Fig. 3d) it is easier to classify "normal" against "abnormal" network traffic in networks with medium mobility (pause time equal to 200 or 400 sec) compared to stationary networks (pause time equal to 700 sec).

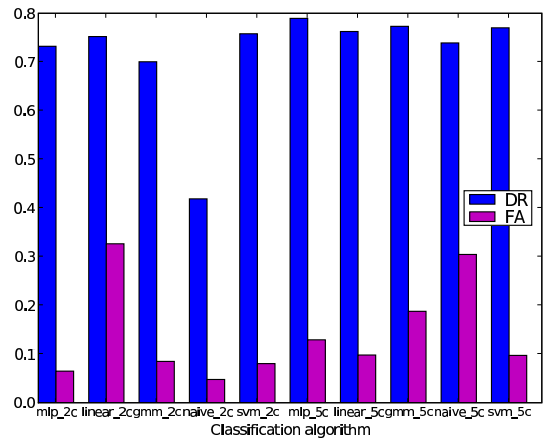

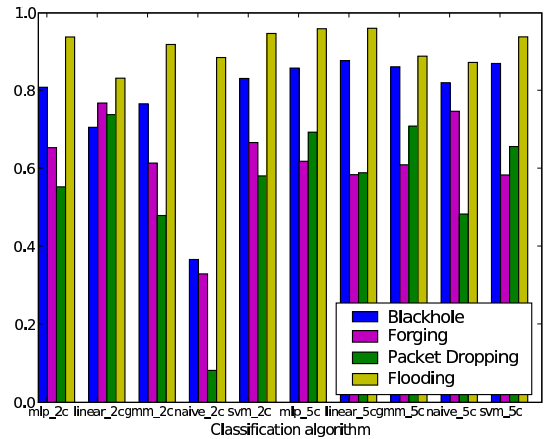

(b) Detection Rate of each type of attacks

**Fig. 2** Comparison of all Classification algorithms

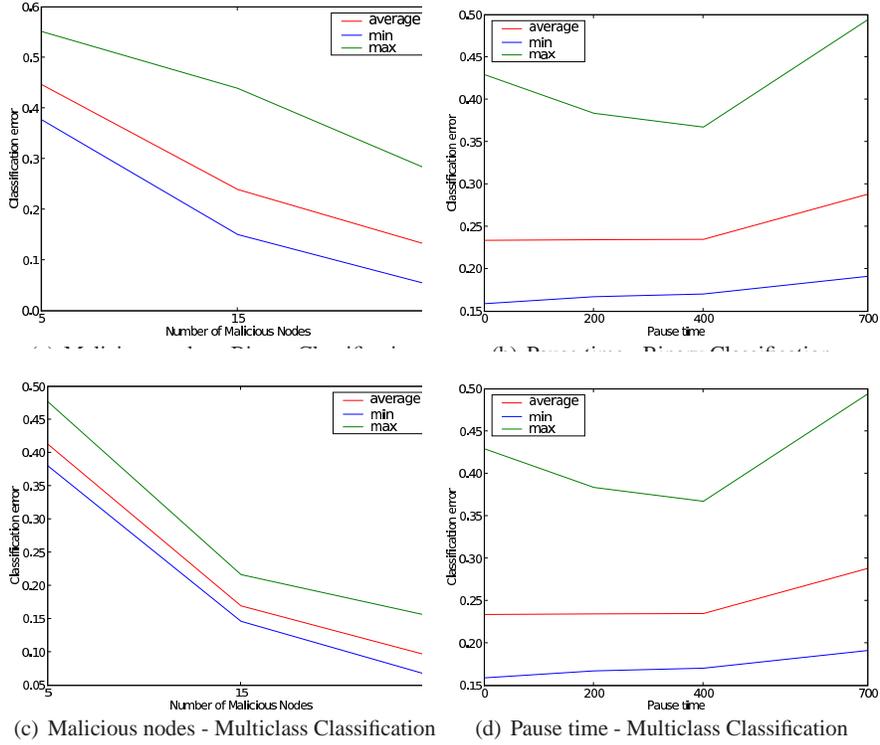

(c) Malicious nodes - Multiclass Classification  (d) Pause time - Multiclass Classification

**Fig. 3** Classification error versus malicious nodes and pause time for binary and multiclass classification

## 4 Conclusions

In this paper we presented a performance comparison of five efficient and commonly used classification algorithms (MultiLayer Perceptron (MLP), Linear classifier, Gaussian Mixture Model (GMM), Naïve Bayes classifier and Support Vector Machines (SVM)) applied to intrusion detection in MANETs. We have used features from the network layer and evaluated the performance of these algorithms for the detection of four serious attacks in MANETs, the *Black hole*, *Forging*, *Packet Dropping* and *Flooding* attack. We investigated which is the most appropriate sampling interval time and concluded that the sampling interval of 15 sec is on average the most efficient, based on the performance of the testing datasets, when the k-fold cross validation of the training datasets is performed randomly.

Furthermore, we concluded that the most efficient classifier for detecting all four types of attacks simultaneously is the SVM classifier for multiclass classification although the MLP classifier presents a satisfying *Detection Rate* (DR) and also a quite high *False Alarm* (FA) rate. The easiest attack to be detected is the *Flooding* attack, while the most difficult attack to detect is the *Packet Dropping* attack, something that was also implied in our previous work [9]. We also investigated how the number of malicious nodes in the network and the mobility of the network affects the performance of the classification algorithms in detecting intrusions. We concluded that the highest the number of malicious node in the network the easiest to detect intrusions. Furthermore, the classification algorithms present effective detection of attacks in MANETs with medium mobility.

For future work, we plan to investigate if the tuning of classification models using non-random k-fold but sequential may give us better performance. Furthermore, we plan to examine if the combination of classifiers and the creation of an ensemble classifier can give us better results.